\documentclass[aps,prd,preprint,superscriptaddress,tightenlines,nofootinbib]{revtex4}

\usepackage{graphicx}
\usepackage{dcolumn}
\usepackage{bm}

\def\Y{\ifmmode \Upsilon \else%
$\Upsilon$%
\fi}
\def\chib{\ifmmode \chi_b \else%
$\chi_b$%
\fi}
\def\chibp{\ifmmode \chi_b' \else%
$\chi_b'$%
\fi}
\def\Q#1#2#3#4{\ifmmode
 \,#1\,{^{#2}#3}_{#4}
\else%
$#1\,{^{#2}#3}_{#4}$ %
\fi}
\def\eonem#1#2{\ifmmode
\left| <#2|r|#1> \right|
\else%
$\left| <#1|r|#2> \right|$
\fi}

\def\ee{\ifmmode e^+e^- \else $e^+e^-$  \fi}
\def\mm{\ifmmode \mu^+\mu^- \else $\mu^+\mu^-$  \fi}
\def\LL{\ifmmode \ell^+\ell^- \else $\ell^+\ell^-$  \fi}

\def\etal{{\it et al.}}
\def\B{{\cal B}}
\def\BR{\B}
\def\br{\B}
\def\a{a^0_1}
\def\nowhitespace{\quad\\[-0.3cm]\quad\\}
\def\nowhitespace{}
\def\capfix{\\[-0.3cm]}

\begin{document}

\preprint{CLNS 08/2033}       
\preprint{CLEO 08-16}         

\title{Search for Light CP-odd Higgs in Radiative Decays of \boldmath{\Y(1S)}}


\author{W.~Love}
\author{V.~Savinov}
\affiliation{University of Pittsburgh, Pittsburgh, Pennsylvania 15260, USA}
\author{H.~Mendez}
\affiliation{University of Puerto Rico, Mayaguez, Puerto Rico 00681}
\author{J.~Y.~Ge}
\author{D.~H.~Miller}
\author{I.~P.~J.~Shipsey}
\author{B.~Xin}
\affiliation{Purdue University, West Lafayette, Indiana 47907, USA}
\author{G.~S.~Adams}
\author{M.~Anderson}
\author{J.~P.~Cummings}
\author{I.~Danko}
\author{D.~Hu}
\author{B.~Moziak}
\author{J.~Napolitano}
\affiliation{Rensselaer Polytechnic Institute, Troy, New York 12180, USA}
% ...
\collaboration{CLEO Collaboration}
\noaffiliation


\date{July 9, 2008} 

\begin{abstract}
We search for a non-SM-like CP-odd Higgs boson ($a^0_1$)
with $m_{a^0_1}< 2 m_b$ in radiative decays of the $\Upsilon(1S)$,
using 21.5M $\Upsilon(1S)$ mesons directly produced in $e^+e^-$ annihilation.
We investigate $a^0_1\to\tau^+\tau^-$ and $a^0_1\to\mu^+\mu^-$ decay channels.
No significant signal is found.
We obtain upper limits on the product of
${\cal B}(\Upsilon(1S)\to\gamma a^0_1)$ and ${\cal B}(a^0_1\to\tau^+\tau^-)$ or
${\cal B}(a^0_1\to\mu^+\mu^-)$.
Our $\tau^+\tau^-$ results are almost two orders of magnitude more 
stringent than previous upper limits.
Our data provide no evidence for a Higgs state 
with a mass of 214 MeV
decaying to $\mu^+\mu^-$.
Existence of such a state was previously proposed as an explanation 
for 3 $\Sigma^+\to p\mu^+\mu^-$ events, having $\mu^+\mu^-$ masses just 
above the kinematic threshold, observed by the HyperCP experiment.   
Our results constrain NMSSM models. 
\end{abstract}

\maketitle

Direct searches at LEP for the Standard Model Higgs boson, a CP-even scalar, set a lower bound on its mass 
in excess of $10^2$ GeV \cite{LEP}.
Many extensions of the Standard Model predict the existence 
of a CP-odd pseudoscalar Higgs boson (hereafter denoted as $\a$), 
which could be light.
For example, the Next-to-Minimal Super-Symmetric Model (NMSSM) 
with $\a$ mass below the threshold for $\a\to b\bar b$ decay is particularly
well motivated \cite{Dermisek}.
Radiative production in $\Y(1S)$ decays, $\Y(1S)\to\gamma\a$, offers a unique 
experimental opportunity to search for such a state.
The couplings of the Higgs to fermions are proportional to their
masses, therefore enhanced with respect to lighter mesons.
The expected rate is given by \cite{ManganoNason}:
$$
 \frac{\br(\Y(1S)\to\gamma\a)}{\br(\Y(1S)\to\mu^+\mu^-)} = \frac{G_F {m_b}^2}{\sqrt{2}\pi\alpha}
  {g_d}^2 \, \left\{ 1 - \left( \frac{m_{\a}}{m_{\Y(1S)}} \right)^2 \right\} \, {\cal C} ,
$$
where $G_F$ is the Fermi constant, $\alpha$ is the fine structure constant, 
$g_d$ is the $\a$ coupling to the down-type
fermions, and ${\cal C}$ incorporates QCD and relativistic corrections.
The coupling $g_d\propto\tan\beta\cos\theta_A$, where $\tan\beta$ is the ratio of vacuum expectations for the
two Higgs doublets, and $\theta_A$ is the mixing angle between doublet and singlet CP-odd Higgs
bosons (for $\theta_A=90^0$, $\a$ is a pure singlet and decouples from fermions);
$g_d$ depends on the detailed choice of SUSY parameters. 
For $m_{\a}<2 m_b$, the decay $\a\to\tau^+\tau^-$ is expected to dominate,
especially at large $\tan\beta$, 
$\BR(\a\to\tau^+\tau)\sim0.9$ \cite{Dermisek}.
For $m_{\a}<2 m_{\tau}$, $\a\to\mu^+\mu^-$ decays are copious below the $s\bar s$ threshold.
In fact, it has been suggested that 
3 $\Sigma^+\to p\mu^+\mu^-$ events observed by the HyperCP experiment \cite{HyperCP}
are due to a CP-odd Higgs with a mass of $214.3\pm0.5$ MeV \cite{HyperCPHiggs}.

The data for this search 
were acquired with the CLEO-III detector \cite{CLEOdet} 
operating at the Cornell Electron
Storage Ring (CESR) and correspond to an integrated luminosity 
of 1.1 fb$^{-1}$ at the $\Y(1S)$, yielding
$(21.5\pm0.4)\times10^6$ resonant decays.
We also use 7 fb$^{-1}$ of data collected at and near the $\Y(4S)$ resonance
for continuum background studies. 

The CLEO-III detector has a solid angle coverage of 93\%\ of $4\pi$
for charged and neutral particles. 
The CsI calorimeter attains $\gamma$ 
energy resolutions of about 2\%\ for $E_\gamma \ge 1$ GeV and  5\%\ at 100 MeV. 
The charged particle tracking system operated in a 1.5 T
magnetic field along the beam axis and achieved a momentum ($p$)
resolution of 0.35\%\ at 1 GeV and a $dE/dx$ resolution of 6\%. 
The muon detector consists of the wire chambers 
located at 3, 5, and 7 hadronic interaction lengths (at normal incidence)
of iron absorber. 

We select events with exactly two tracks of opposite charge and at least
one $\gamma$. The highest energy $\gamma$, which when paired with any other
$\gamma$ in the event, is not within three standard deviations ($3\sigma$) of
the $\pi^0$ mass,
is selected to be a candidate for $\Upsilon(1S)\to\gamma{\a}$.
Photons are defined as showers that do not match charged track projections into the calorimeter.
For optimal energy resolution and smallest backgrounds
the radiative-decay $\gamma$ is required to be in the barrel part of the
calorimeter ($|\cos\theta|<0.8$). 
Its lateral shower profile must be consistent with an
isolated electromagnetic shower.
When applying the $\pi^0$ veto, we place loose requirements on the other 
(non-signal) photon. Namely, the other $\gamma$ is not required to be isolated
and the entire solid angle of the calorimeter is used for its detection.
Its energy is required to be at least $30$ MeV ($60$ MeV)
in the barrel (endcap) part of the calorimeter.
The $\pi^0$ veto suppresses $e^+e^-\to\tau^+\tau^-$,
with at least one $\tau$ decaying to a channel with $\pi^0$ (mostly $\rho\nu$).

To select $\a\to\tau^+\tau^-$ candidates we require a missing energy between 2 and 7 GeV.
The total energy calculation is based on charged track momenta (the pion mass is assumed)
and calorimeter energy for the looser definition of the $\gamma$ candidates (see above).
To suppress hadronic events from the continuum production and $\Upsilon(1S)$ decays, 
at least one of the charged tracks must be identified as an electron or a muon.
Events with two electrons are discarded to suppress Bhabha like events.
Besides being identified in the muon system, 
the $\mu$ candidate is required to have 
an energy deposited in the calorimeter ($E$) consistent with a 
minimum ionizing particle. 
The electron candidate must have $E$ equal $p$ within $\pm15\%$.
$dE/dx$ consistency is required for for leptons.
The invariant mass of photons (except for the radiative-decay $\gamma$) plus the charged track 
not identified as a lepton is required to be less than $2$ GeV.
To suppress final state radiation, the cosine of the angle between any charged track
and the $\gamma$ candidate must be less than $0.99$.

We search for evidence for a signal of a monochromatic peak
in the $\gamma$ energy distribution. 
Thus, our results assume that the $\a$ natural width is negligible compared 
with the experimental resolution, an assumption which is expected to be true, with 
the exception of the heaviest masses probed in the $\tau^+\tau^-$ channel, 
due to possible mixing with the $\eta_b$ \cite{etabmixing}.
CLEO previously published an alternative method for 
probing the $\a$ mass approaching the $b\bar b$ states, which is not 
sensitive to assumptions about its width \cite{Duboscq}.

\begin{figure}[tbhp]
\nowhitespace
\includegraphics[width=\hsize]{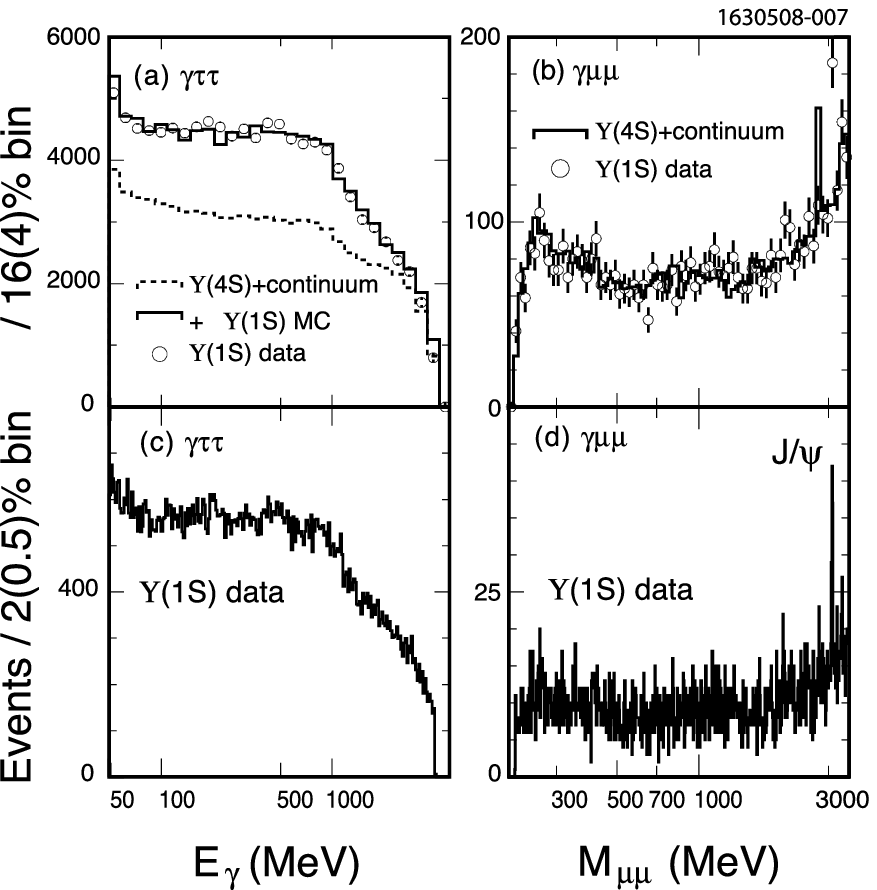} 
\capfix
\caption{
\label{fig:data} 
Photon energy and dimuon mass distributions in $\gamma\tau^+\tau^-$ (a,c) and 
$\gamma\mu^+\mu^-$ (b,d) data, respectively.
Bin size for the right column plots is given in the axes labels in parentheses.
In the top row,
the $\Y(1S)$ data (points with error bars) are 
compared to the estimated backgrounds (dashed and solid lines).
In the bottom row, the $\Y(1S)$ data (solid line) are shown in fine
binning comparable to the detector resolution (see bottom row of Fig.~\ref{fig:llmc}).
In (b) the $J/\psi$ ISR peak is shifted in the background estimate, since
we scaled $\mu$ momenta down by the ratio of the beam energies when 
scaling the higher energy data to the $\Y(1S)$ distribution.
}
\end{figure}

The selected event sample is composed mostly of continuum $e^+e^-\to (\gamma)\tau^+\tau^-$ events, 
where the $\gamma$ candidate 
comes either from initial state radiation (ISR) or from a $\pi^0$ produced in $\tau$ decay, 
with the second $\gamma$ not reconstructed.
The background estimates are superimposed on top of the spectrum obtained 
at the $\Y(1S)$ resonance in Fig.~\ref{fig:data}a.
The continuum backgrounds are estimated by scaling the $\Y(4S)$ distributions.
There is also a significant contribution from $\Y(1S)\to\tau^+\tau^-$ with the $\gamma$ candidate 
coming from a $\pi^0$ decay. 
The observed $\gamma$ spectrum with binning comparable to our $\gamma$ energy 
resolution is shown in Fig.~\ref{fig:data}c.
No significant peaks are observed.

\begin{figure}[bthp]
\nowhitespace
\includegraphics[width=\hsize]{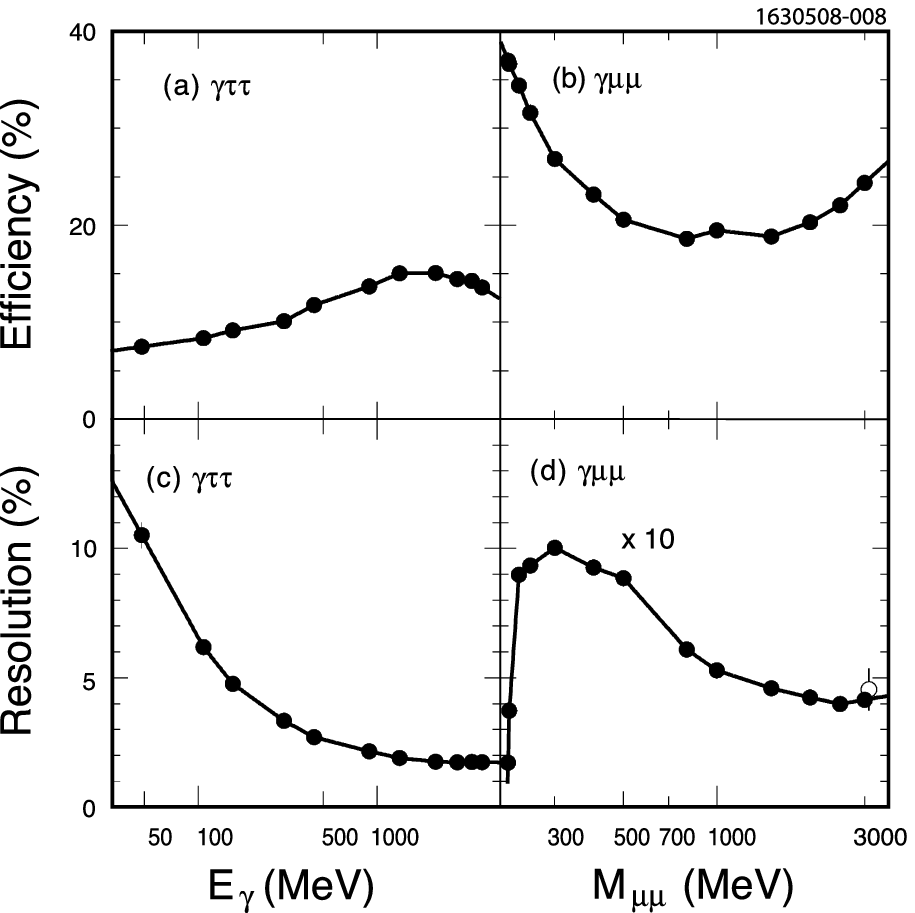} 
\capfix
\caption{
\label{fig:llmc} 
Efficiency (a,b) and 
$\a$ mass resolution (c,d)
obtained from the fits to the 
$\a\to\tau^+\tau^-$ (left column) and 
$\a\to\mu^+\mu^-$ (right column) 
signal MC (points)
and interpolated for the regions in between (solid line).
In (d) relative dimuon mass resolution was multiplied by a factor of 10.
See Appendix C of Ref.~\cite{ManganoNason} for 
explanation of improvement of the dimuon mass
resolution near the kinematic threshold.  
The hollow point with the error bar in (d) represents the fit of the mass resolution
to the $J/\psi\to\mu^+\mu^-$ ISR peak observed in the $\Y(1S)$ data.
}
\end{figure}

The channel $\a\to \mu^{+}\mu^{-}$ is selected by
identifying both muons.
We require that the total observed energy be within 250 MeV of 
the center-of-mass energy.
The invariant dimuon mass has better resolution than the $\gamma$ energy, therefore
we use it to look for the $\a$ signal.
The selected data are dominated by radiative $\mu$-pairs with a hard radiative
photon. 
The data selected at the peak of the $\Y(1S)$ resonance 
are well described by scaling 
the data collected at and near the $\Y(4S)$ 
as illustrated in Fig.~\ref{fig:data}b.
The $\Y(1S)$ distribution plotted with binning comparable to our dimuon mass resolution 
is shown in Fig.~\ref{fig:data}d.
No significant peak is found except for the $J/\psi$ produced by ISR.

The signal efficiency varies with Higgs mass, or equivalently, $\gamma$ energy.
In order to determine the efficiency,
we generated signal Monte Carlo (MC) for several $\a$ masses and interpolated
for masses in between.
Proper angular correlations were implemented in the MC 
for both the polar angle of the radiative $\gamma$ and for $\tau$ 
polarizations \cite{Was}.
The signal peaks observed in the dimuon mass ($\gamma$ energy) distribution  
for $\a\to\mu^+\mu^-$ ($\a\to\tau^+\tau^-$)
were fitted to a Gaussian (with an asymmetric low energy tail, i.e., a Crystal Ball
line shape \cite{CBLineShape}) to determine reconstruction efficiency and detector resolution, which are
shown in Fig.~\ref{fig:llmc}ac (bd).
The fit to the $J/\psi\to\mu^+\mu^-$ ISR peak observed in the $\Y(1S)$ data 
gives a value for the resolution consistent with the MC expectations for an $\a\to\mu^+\mu^-$ 
signal at that mass (Fig.~\ref{fig:llmc}d).

We have scanned the observed $\gamma$ energy and dimuon mass distributions
by fitting a signal peak on top 
of a cubic background polynomial, 
changing the peak position in steps equal to the detector resolution.
The peak width was fixed to the MC expectations.
The fit range was set to $\pm0.5$ of $\ln(E_\gamma)$ 
($\pm0.25$ of $\ln(M_{\mu\mu})$) around the peak position.
The $J/\psi\to\mu^+\mu^-$ peak region was excluded within $\pm3\sigma$, unless fitting a peak 
at the $J/\psi$ mass.
Since, in the dimuon channel, the continuum backgrounds 
saturate the $\Y(1S)$ sample (see Fig.~\ref{fig:data}b),
we simultaneously fit the background polynomial to the 
dimuon mass distribution obtained from the higher statistics $\Y(4S)$ data.
The overall background normalization factor for the $\Y(4S)$ data was 
included as a free parameter in the fit.

To test for possible bias in the fit procedure, we calculated
average and root mean square (RMS) values of fitted signal amplitude ($N$) divided by its error ($\Delta N$).
In absence of any signal peaks, values of 0.0 and 1.0 are expected, respectively.
The average value for the data
is $+0.01\pm0.09$ ($-0.06\pm0.05$) for the $\gamma\tau^+\tau^-$ 
($\gamma\mu^+\mu^-$) sample, while
the RMS is $1.16\pm0.09$ ($1.05\pm0.05$).
To cover the observed deviations from the expectations
we assume 15\%\ for a possible systematic error in the fit procedure.

The ratio of the fit likelihoods for
a signal peak included in the fit (${\cal L}_{max}$) 
and the data fit with the background term alone (${\cal L}_{0}$) is used to 
calculate the peak significance in standard deviations, 
$\sqrt{2\ln({\cal L}_{max}/{\cal L}_{0})}$.
No peak with significance above $3\sigma$ is found in the $\gamma\tau^+\tau^-$ data.
In the $\gamma\mu^+\mu^-$ data, the ISR $J/\psi$ peak has $8.3\sigma$ significance.
There are two other mass points which produce peaks with significance slightly above $3\sigma$:
$3.3\sigma$ at $2041\pm4$ MeV 
and $3.1\sigma$ at $211.92\pm0.15$ MeV.
For one trial, the probability ($\epsilon$) of  background fluctuations producing a peak  
with significance equal or larger to $3.3\sigma$ ($3.1\sigma$)
is  $0.05\%$ ($0.1\%$). 
We performed $482$ fits to $\gamma\mu^+\mu^-$ spectrum with the peak positions 
separated by one unit of mass resolution.
Assuming that peaks must be separated by at least 3 units of mass resolution to fluctuate 
independently, we performed about $n=482/3\approx 160$ statistically independent trials. 
The overall probability in our scan of producing  
at least one peak with significance of at least $3.3\sigma$ ($3.1\sigma$) is
$1 - (1-\epsilon)^n \approx 8\%$ ($15\%$).

\begin{figure}[tbhp]
\nowhitespace
\includegraphics[width=\hsize]{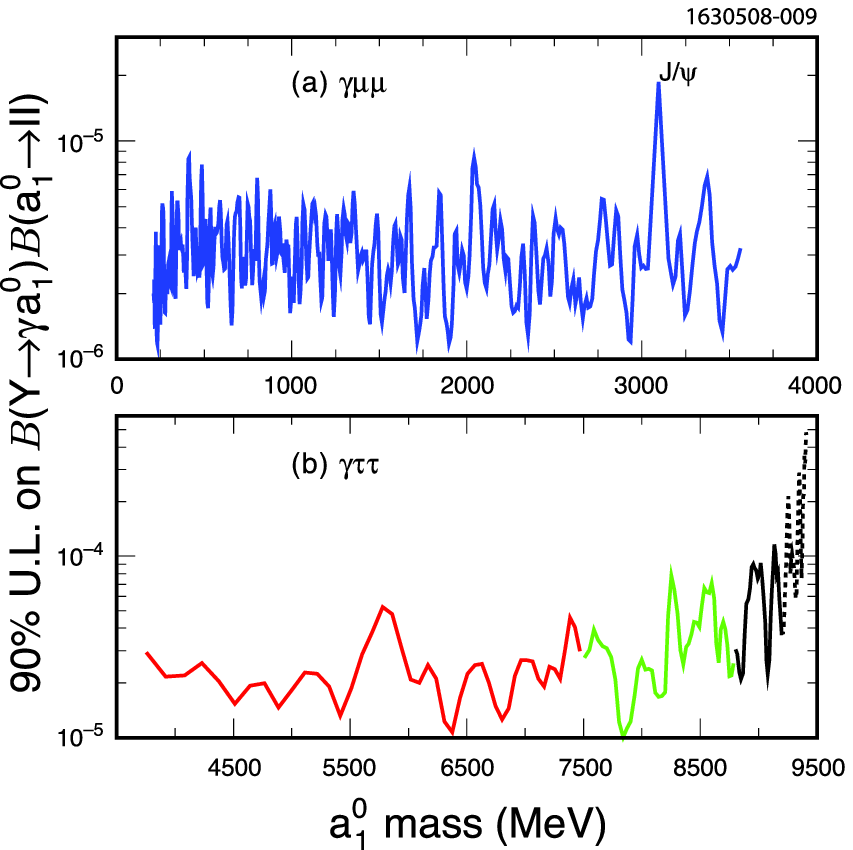} 
\capfix
\caption{
\label{fig:ul}
Upper limits on $\BR(\Y(1S)\to\gamma\a)$ (a) $\times\BR(\a\to \mu^+\mu^-)$ 
(b) $\times\BR(\a\to \tau^+\tau^-)$  as a function of the $\a$ mass (90\%\ C.L.).
The color coding corresponds to the one used in Fig.~\ref{fig:nmssm}.
The dashed line indicates the region ($m_{\a}>9.2$ GeV) where $\a$ is likely to mix with $\eta_b$ and
acquire a non-negligible width, thus invalidating our analysis method.
}
\end{figure}

\begin{figure}[bthp]
\includegraphics[width=\hsize]{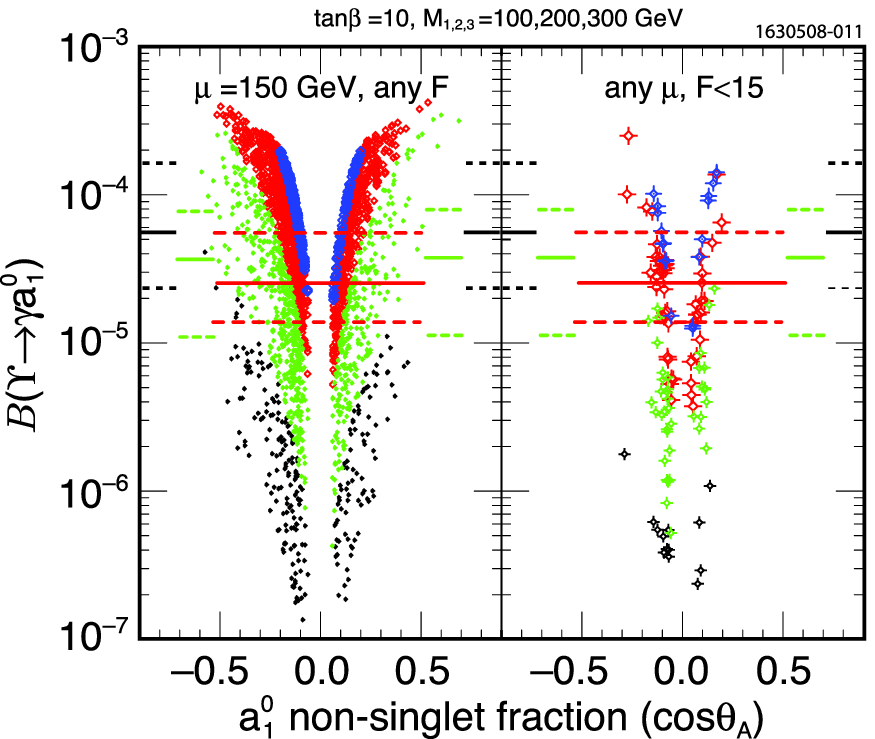} 
\caption{\label{fig:nmssm} 
Comparison of CLEO upper limits on $\BR(\Y(1S)\to\gamma\a)\times\BR(\a\to\tau^+\tau^-)$
(solid and dashed lines) 
to the NMSSM predictions by 
Dermisek, Gunion, McElrath (points) \cite{Dermisek}.
See the text for explanations.}  
\end{figure}

With no evidence for an $\a$ signal we set upper limits on its possible 
production rate as a function of the $\a$ mass.
To determine upper limits on the signal event yield we
fix the signal amplitude to positive values and minimize it with respect to the
background parameters to obtain a likelihood for given signal amplitude.
We then integrate the likelihood function and find a signal amplitude which
bounds 90\%\ of the total area.
Dividing this event limit by efficiency and the number of $\Y(1S)$ decays in our data we
obtain upper limits on $\BR(\Y(1S)\to\gamma\a)\times\BR(\a\to l^+l^-)$ ($l=\tau$ or $\mu$),
which are displayed in Fig.~\ref{fig:ul}.
The limits were scaled up by 20\%\ to account for the possible systematic error in the fit procedure (15\%),
in the efficiency calculation (MC statistics, interpolation between MC points,
detector modeling; together $<10\%$), in number of $\Y(1S)$ decays (2\%), and
in simulation of detector resolution (10\%).
Our upper limits on $\BR(\Y(1S)\to\gamma\a)\times\BR(\a\to\tau^+\tau^-)$
are almost 2 orders of magnitude more stringent than previously obtained by ARGUS \cite{ARGUS}.
Our upper limits on $\BR(\Y(1S)\to\gamma\a)\times\BR(\a\to\mu^+\mu^-)$ are the first 
experimental bounds.

Our $\gamma\tau^+\tau^-$ results provide
stringent constraints on NMSSM models, eliminating a large portion of
previously unconstrained parameter space.
This is illustrated in Fig.~\ref{fig:nmssm}, where 
NMSSM calculations by Dermisek, Gunion, McElrath \cite{Dermisek} 
are compared to our 
upper limits. 
While some model parameters were fixed (e.g. $\tan\beta=10$) in the theoretical
calculations, other parameters were sampled.
Each point represents a different choice of NMSSM parameter values
consistent with the current experimental constraints.
The plot on the right represents models with the additional requirement of low
fine-tuning of electroweak symmetry breaking (see Ref.\cite{Dermisek} 
 for details). Color coding corresponds to various $\a$ mass ranges.
Our upper limits in various $\a$ mass ranges are shown by horizontal lines.
Solid (dashed) line(s) represent an average (minimal and maximal) 
upper limits in given mass range. 
Assuming that $\BR(\a\to \tau^+\tau^-)$ is 100\%, 
the models above these lines are excluded.
Our upper limits in the $\gamma\tau^+\tau^-$ channel for lower $\a$ masses,  
$2 m_{\tau} <m_{\a}<7.5$ GeV (red lines at the center of each plot), 
eliminate a significant fraction of models in 
this mass range (red points). Only very few models are challenged in the
$7.5<m_{\a}<8.8$ GeV range (green points and lines at the sides of scatter plots). 
For higher masses (black points and lines touching the plot axes)
our discriminating power fades away as the backgrounds increase while the expected signal rate
decreases due to the phase space suppression.
The non-singlet fraction of $\a$ ($\cos\theta_A$) increases with falling $\tan\beta$ \cite{Dermisek},
though the net effect on $g_d$, and therefore $\BR(\Y(1S)\to\gamma\a)$, is to decrease.
At the same time, $\a$ coupling to up-type fermions, $g_u\propto\cos\theta_A/\tan\beta$,
increases, lowering $\BR(\a\to \tau^+\tau^-)$.
Thus, models with small $\tan\beta$ values are less constrained by our data.

\begin{figure}[tbhp]
\nowhitespace
\includegraphics[width=\hsize]{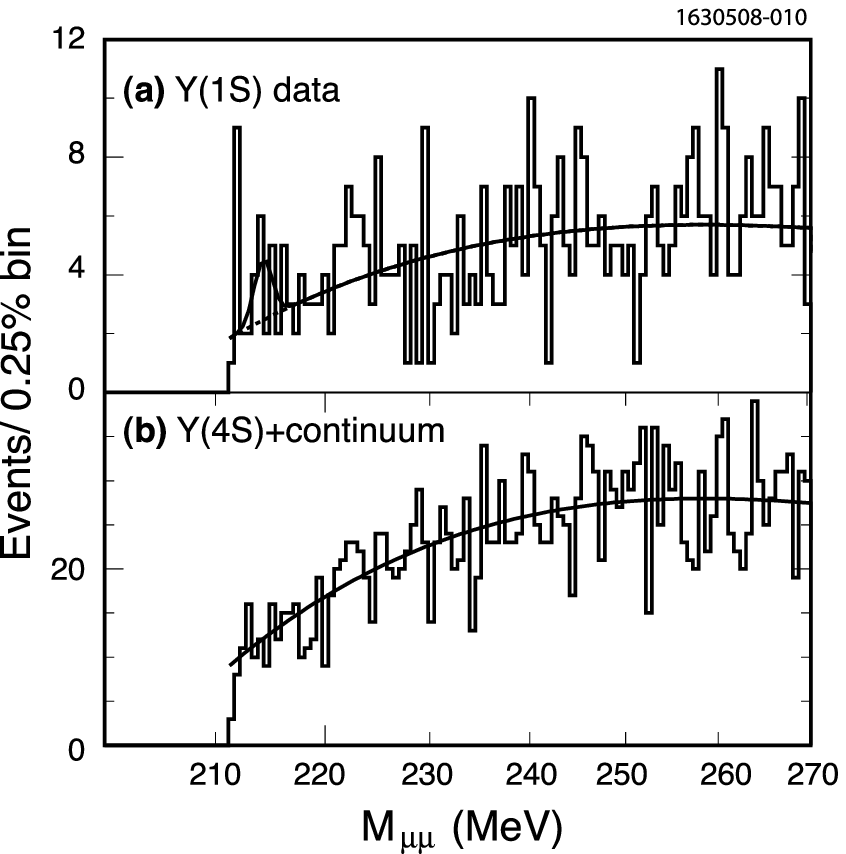} 
\capfix
\caption{\label{fig:hypercp}
 A fit of a peak at a dimuon mass of 214.3 MeV with fixed width at the expected mass
resolution, on top of cubic polynomial to our
$\gamma\mu^+\mu^-$ data obtained at the $\Y(1S)$ center-of-mass energy (a).
The polynomial describing the continuum backgrounds was simultaneously constrained 
by the data collected at and near the $\Y(4S)$ resonance (b).
}
\end{figure}

Since $\BR(\a\to\mu^+\mu^-)$ is not expected to be large above the $\a\to s\bar s$ 
threshold ($\sim1$ GeV), we did not transfer our limits on 
$\BR(\Y(1S)\to\gamma\a)\BR(\a\to \mu^+\mu^-)$ to  
Fig.~\ref{fig:nmssm}, where NSSM model calculations of $\BR(\Y(1S)\to\gamma\a)$
below the $\tau^+\tau^-$ threshold (blue points)
were performed only for $m_{\a}>1$ GeV. 
Our limits below this mass value constrain NSSM scenarions.
Of particular interest is our upper limit 
for $m_{\a}=214.3$ MeV, i.e., the $\mu^+\mu^-$ mass of 3 $\Sigma^+\to p\mu^+\mu^-$
events observed by the HyperCP experiment \cite{HyperCP}. 
The fit to our data (see Fig.~\ref{fig:hypercp}) gives $7.0^{+5.3}_{-4.5}$ events at this mass
and leads to an upper limit
of $\BR(\Y(1S)\to\gamma\a)< 2.3\times 10^{-6}$ at 90\% C.L.
He, Tandean and Valencia 
showed that they could explain the HyperCP events with the $\a$ hypothesis
and still be consistent with the 
constraints from $K\to \pi\mu^+\mu^-$ experiments \cite{HyperCPHiggs}.
In their calculations they used $g_d^2=0.12$, while
our upper limit translates into $g_d^2<0.026$ (using ${\cal C}=0.5$ \cite{ManganoNason}), 
which calls for reevaluation of
the $\a$ hypothesis for the HyperCP events.

In summary,
we have obtained meaningful upper limits on
$\br(\Upsilon(1S)\rightarrow \gamma \a)\times \br(\a\rightarrow
\tau^{+}\tau^{-})$ and $\br(\Upsilon(1S)\rightarrow \gamma
\a)\times\br(\a\rightarrow \mu^{+}\mu^{-})$. Our limits on
$\gamma\tau^{+}\tau^{-}$ are almost two orders of magnitude more
stringent than those from ARGUS and eliminate a large portion of
previously unconstrained parameter space in NMSSM models.
Our limits on $\gamma\mu^{+}\mu^{-}$ challenge models with $\a$ mass below
$s\bar s$ threshold and the $\a$ interpretation of HyperCP 
$\Sigma^+\to p\mu^+\mu^-$ events.
Our limits are applicable to any light scalar or pseduo-scalar boson, 
which arises in various extensions of Standard Model.   

We gratefully acknowledge the effort of the CESR staff 
in providing us with excellent luminosity and running conditions. 
This work was supported by 
the A.P.~Sloan Foundation, 
the National Science Foundation, 
the U.S. Department of Energy, 
the Natural Sciences and Engineering Research Council of Canada, and 
the U.K. Science and Technology Facilities Council.
We thank Radovan Dermisek,  Jack Gunion, Michelangelo Mangano, German Valencia and  Miquel Sanchis-Lozano 
for useful discussions of NMSSM models.

\end{document}